\documentclass[preprint,5p,times,twocolumn]{elsarticle}

\usepackage{amssymb}
\usepackage{amsmath}
\usepackage{setspace}
 
\usepackage{graphicx} 
\usepackage{color}
\usepackage{makecell}

\journal{Computational Condensed Matter}

\begin{document}

\begin{frontmatter}

\title{Compressed magnetic Moment Tensor Potentials via low-rank matrix and tensor factorizations}

\author[Skoltech,HSE]{Alexey S. Kotykhov\corref{cor1}}
\ead{Alexey.Kotykhov@skoltech.ru}
\cortext[cor1]{Corresponding author.}

\author[Skoltech,DM]{Alexander V. Shapeev}

\author[HSE]{Maxim Rakhuba}

\author[HSE]{Ivan S. Novikov}

\affiliation[Skoltech]{organization={Skolkovo Institute of Science and Technology, Skolkovo Innovation Center}, 
addressline={Bolshoy boulevard 30}, 
city={Moscow},
postcode={121205},
country={Russian Federation}}

\affiliation[HSE]{organization={HSE University, Faculty of Computer Science}, 
addressline={Pokrovsky boulevard 11}, 
city={Moscow},
postcode={109028},
country={Russian Federation}}

\affiliation[DM]{organization={Digital Materials LLC}, 
addressline={Odintsovo, Kutuzovskaya str. 4A}, 
city={Moscow Region},
postcode={143001},
country={Russian Federation}}
    
\date{August 2026}

\begin{abstract}

We propose a parameter-reduced version of magnetic Moment Tensor Potential (mMTP) based on various matrix and tensor decompositions. The resulting compressed magnetic machine-learning potential reduces the number of parameters by a factor of approximately $1.5-3$ without compromising predictive performance on the validation set. We evaluate the performance of the compressed potentials for magnetic, structural, and vibrational properties, as well as in molecular dynamics simulations of Fe–Al and CrN systems. We demonstrate that the simulation results obtained with the compressed mMTP are numerically consistent with those of the original uncompressed model and are in agreement with density functional theory calculations and experimental data.

\end{abstract}

\end{frontmatter}

\section{Introduction}

Density functional theory (DFT) calculations are among the most widely used approaches in computational materials science \cite{maurer2019advances}. However, DFT is computationally expensive and is often limited to systems containing a few hundred atoms. To overcome this limitation, many machine-learning interatomic potentials (MLIPs) \cite{behler2007generalized, bartok2010gaussian, thompson2015spectral, shapeev2016moment, drautz2019atomic, batzner20223, batatia2022mace, chen2022universal} have been proposed to approximate DFT data at a substantially lower computational cost.

An explicit treatment of magnetism is important for accurately describing magnetic materials \cite{herper1999ab, friak2010ab, kormann2012atomic, kormann2016influence, song2021negative}. Therefore, it is necessary to perform magnetism-inclusive modeling, for example using spin-polarized DFT calculations, which are even more computationally expensive than non-spin-polarized calculations. Several MLIPs have therefore been developed to describe magnetic materials. For example, magnetism can be incorporated by introducing distinct atom types for up and down spins \cite{shenoy2024collinear, bienvenu2025development}, or by combining the Heisenberg model with MLIPs \cite{nikolov2021data, chapman2022machine}. Other models explicitly incorporate magnetic degrees of freedom into their functional forms, including neural network potentials \cite{eckhoff2021high, chapman2022machine, yang2024deep}, SNAP \cite{domina2022spectral}, MTP \cite{novikov2022magnetic}, GAP \cite{gao2024machine}, ACE \cite{rinaldi2024non}, graph neural networks \cite{yu2024spin}, and equivariant neural networks \cite{yuan2024equivariant}. An advantage of magnetic MLIPs is that they enable Monte Carlo simulations for the magnetic degrees of freedom \cite{eckhoff2021high, rinaldi2024non, bienvenu2025development} and spin-lattice dynamics \cite{chapman2022machine, yu2024spin}, which are computationally expensive to perform using DFT.

The challenge of large numbers of trainable parameters has also emerged in other areas of machine learning, including convolutional neural networks and large language models. In these fields, low-rank tensor decompositions have been successfully used to significantly reduce the number of trainable parameters with only a slight decrease in accuracy \cite{deng2020model, hu2022lora, li2023model}. Similar approaches have been explored for nonmagnetic MLIPs, where tensor decompositions were used to compress nonmagnetic MTP and ACE models \cite{vorotnikov2025low}. Since magnetic MLIPs typically contain even more trainable parameters than their nonmagnetic counterparts and may therefore require larger training sets, parameter compression is particularly important for magnetic models.

In this study, we focus on the compression of the magnetic Moment Tensor Potential (mMTP) \cite{novikov2022magnetic}, which has at least four times as many radial parameters as its non-magnetic version. To reduce the number of mMTP parameters, we utilize various tensor decompositions. Models employing tensor decompositions will henceforth be referred to as compressed potentials. We evaluate the developed compressed potentials for the Fe--Al and CrN systems and demonstrate that the number of parameters can be reduced by a factor of $1.5$ -- $3$ without compromising predictive accuracy. Furthermore, the structural, magnetic, and vibrational properties obtained with the compressed potential correspond to those of the uncompressed potential and are consistent with DFT and experimental data.

\section{Methodology}

\subsection{Magnetic Moment Tensor Potential}

Magnetic Moment Tensor Potential is a generalization of the non-magnetic MTP to the case of collinear magnetism. The functional form of mMTP is described in more detail in previous works \cite{novikov2022magnetic, kotykhov2023constrained}; here, we present the basic formulas and concepts. 

The atomic local environment $\mathfrak{n}_i$ of atom $i$ contains information about the type $z_i$ and magnetic moment $m_i$ of the central atom, as well as the positions of the neighbors $\textbf{r}_{ij}$ relative to the central atom, atomic types $z_j$, and magnetic moments $m_j$ of neighboring atoms, which are defined as those located within the cutoff radius $R_{\text{cut}}$ from the central atom. 

The energy of a system of $N$ atoms is decomposed into a sum of the energies of local environments $V$:

\begin{equation}\label{eq:E}
    E = \sum_{i=1}^{N} V(\mathfrak{n}_i).
\end{equation}

The energies of local environments are represented as a linear combination of basis functions $B_{\alpha}$:

\begin{equation}\label{eq:local_E}
    V(\mathfrak{n}_i) = \sum_{\alpha = 1}^{N_{\text{linear}}} \xi_{\alpha} B_{\alpha}(\mathfrak{n}_i) + \chi_{z_i},
\end{equation}
where the linear parameters $\Xi = \{\xi_{\alpha}\}$ and energy-shift parameters $X = \{ \chi_{z_i} \}$ are trainable. The number of basis functions $N_{\text{linear}}$ is determined by the level of the mMTP potential. The energy-shift coefficient $\chi_{z_i}$ is defined separately for each atomic type. Therefore, for a system containing $N_{\text{spec}}$ atomic species, there are $N_{\text{spec}}$ coefficients of this type. 

The basis functions consist of various scalar-producing tensor contractions of moments $M_{\mu, \nu}$. The moments themselves consist of a radial $f_{\mu}$ and an angular part $\textbf{r}_{ij}^{\otimes \nu}$:

\begin{equation}\label{eq:moment}
    M_{\mu, \nu}(\mathfrak{n}_i) = \sum_{j=1}^{N_{\text{nbh}}} f_{\mu}(|\textbf{r}_{ij}|, z_i, z_j, m_i, m_j) \cdot \textbf{r}_{ij}^{\otimes \nu},
\end{equation}
where $N_{\text{nbh}}$ denotes the number of neighboring atoms of the $i$-th atom. The symbol ``$\otimes$'' denotes the outer product; therefore, the angular part is an $\nu$-dimensional tensor. 

The radial part is defined as:

\begin{equation}\label{eq:radial}
\displaystyle
\begin{aligned}
    f_{\mu}(|\textbf{r}_{ij}|, z_i, z_j, m_i, m_j) = \sum_{\zeta = 1}^{N_{\text{rad}}} \sum_{\beta = 1}^{N_{\text{mag}}} \sum_{\gamma = 1}^{N_{\text{mag}}} c_{\mu, \zeta, z_i, z_j, \beta, \gamma} \cdot \\ 
    \displaystyle
    \psi_{\zeta}^{\text{rad}}(|\textbf{r}_{ij}|) \psi_{\beta}^{\text{mag}}(m_i) \psi_{\gamma}^{\text{mag}}(m_j) (R_{\text{cut}} - |\textbf{r}_{ij}|)^2.
\end{aligned}
\end{equation}

The functions $\psi^{\text{rad}}$ and $\psi^{\text{mag}}$ are Chebyshev polynomials of the first kind. The numbers $N_{\text{rad}}$ and $N_{\text{mag}}$ specify the sizes of the radial and magnetic bases, respectively, and serve as hyperparameters of the potential. 

As can be seen from formula \eqref{eq:radial}, all magnetism is incorporated into the radial part. Therefore, the potential is capable of describing only collinear magnetism. The tensor $C = \{ c_{\mu, \zeta, z_i, z_j, \beta, \gamma} \}$, containing trainable parameters, is six-dimensional and has dimensions $ N_{\text{rf}} \times N_{\text{rad}} \times N_{\text{spec}} \times N_{\text{spec}} \times N_{\text{mag}} \times N_{\text{mag}}$, where $N_{\text{rf}}$ denotes the number of radial functions $f$ and defines the maximum possible value of $\mu$.  In the non-magnetic case, the tensor $C$ is only four-dimensional and contains at least four times fewer parameters than the corresponding tensor in magnetic MTP.

To restrict the number of possible basis functions, the concepts of the moment level and the basis function level are introduced:

\begin{equation}\label{eq:lelvel_moment}
    \text{lev}(M_{\mu, \nu}) = 2 + 4\mu + \nu,
\end{equation}

\begin{equation}\label{eq:lelvel_basis_func}
    \text{lev}(B_{\alpha}) = \sum_{k} \text{lev}(M_{\mu_{k}, \nu_{k}}).
\end{equation}

The potential includes only those basis functions whose level does not exceed the potential level $\text{lev}_{\text{max}}$, which is a hyperparameter that determines the maximum number $N_{\text{linear}}$ and level of basis functions, thereby controlling the flexibility of the functional form and the maximum order of many-body interactions. The total number of parameters increases with the potential level, as both the number of linear parameters $N_{\text{linear}}$ and the dimension $N_{\text{rf}}$ of the radial-parameter tensor increase. An example of the levels used in this work can be found in Table \ref{tab:params_from_level}.

\begin{table}[!h]
\caption{Number of radial functions $N_{\text{rf}}$ and number of basis functions $N_{\text{linear}}$ for different mMTP levels.}
\label{tab:params_from_level}
\begin{center}
\begin{tabular}{ccc}
\hline
\ level \ & \: $N_{\text{rf}}$ \: & \; $N_{\text{linear}}$ \; \\ \hline
10    & 3     & 16   \\ 
12    & 3     & 29   \\ 
14    & 4     & 52   \\ 
16    & 4     & 92   \\
18    & 5     & 163   \\ \hline
\end{tabular}
\end{center}
\end{table}

In total, the set of all trainable parameters $\Theta$ consists of the tensor of the radial parameters $C$, the linear parameters $\Xi$, and the energy shifts $X$:

\begin{equation}\label{eq:lelvel_moment}
    \Theta = \{ \Xi, X, C \}.
\end{equation}

In this work, the tensor $C$ is compressed to reduce the number of trainable model parameters.

\subsection{Matrix and tensor factorizations for radial parameters}

In this section, we consider methods for factorizing the radial parameter tensor $C$. The axes of the six-dimensional tensor $C \in \mathbb{R}^{N_{\text{rf}} \times N_{\text{rad}} \times N_{\text{spec}} \times N_{\text{spec}} \times N_{\text{mag}} \times N_{\text{mag}}}$ can be permuted, for instance, to obtain $\hat{C}$:
\begin{equation}\label{eq:reshape}
\hat{C} \in \mathbb{R}^{N_{\text{rf}} \times N_{\text{mag}} \times N_{\text{mag}} \times N_{\text{rad}} \times N_{\text{spec}} \times N_{\text{spec}}}. 
\end{equation}
The tensor $\hat{C}$ can then be reshaped into the matrix $\tilde{C} \in \mathbb{R}^{(N_{\text{rf}} N_{\text{mag}}^2) \times (N_{\text{rad}} N_{\text{spec}}^2)}$. Defining $N_{\text{left}} = N_{\text{rf}} N_{\text{mag}}^2$ and $N_{\text{right}} = N_{\text{rad}} N_{\text{spec}}^2$, the reshaping operation can be written as:

\begin{equation}\label{eq:reshape}
    \tilde{C} = \text{reshape}_{(N_{\text{left}}, N_{\text{right}})}(\hat{C}). 
\end{equation}

The matrix $\tilde{C}$ can then be decomposed using matrix factorization:

\begin{equation}\label{eq:mf}
    \tilde{C} \approx U V^{T},
\end{equation}
where $U \in \mathbb{R}^{N_{\text{left}} \times r}$ and $V \in \mathbb{R}^{N_{\text{right}} \times r}$, with $r$ being the chosen rank. Matrix factorization reduces the number of radial parameters from $N_{\text{left}} \cdot N_{\text{right}}$ to $r \cdot (N_{\text{left}} + N_{\text{right}})$ when a sufficiently small rank $r$ is used.

The radial parameter tensor can also be compressed using Tucker decomposition \cite{tucker1966some}:
\begin{equation}\label{eq:Tucker}
\displaystyle
\begin{aligned}
    c_{\mu, \zeta, z_i, z_j, \beta, \gamma} = \sum_{\alpha_1 = 1}^{r_1} \dots \sum_{\alpha_6 = 1}^{r_6} G^{(1)}_{\zeta, \alpha_1} G^{(2)}_{z_i, \alpha_2} G^{(3)}_{\beta, \alpha_3} \\
    \displaystyle
     G^{(4)}_{z_j, \alpha_4} G^{(5)}_{\gamma, \alpha_5} G^{(6)}_{\mu, \alpha_6} S_{\alpha_1, \alpha_2, \alpha_3, \alpha_4, \alpha_5, \alpha_6}, 
\end{aligned}
\end{equation}
where ranks $r_1, \dots, r_6$ are hyperparameters to be chosen, and $G^{(k)}$ are matrices and $S$ is sixth-order tensor. After applying Tucker decomposition to the tensor of the radial parameters, the number of trainable parameters becomes $N_{\text{rad}} r_1 + N_{\text{spec}} r_2 + N_{\text{mag}} r_3 + N_{\text{spec}} r_4 + N_{\text{mag}} r_5 + N_{\text{rf}} r_6 + r_1 r_2 r_3 r_4 r_5 r_6$ instead of $N_{\text{rf}} N_{\text{rad}} N_{\text{spec}}^2 N_{\text{mag}}^2$.

Tensor Train (TT) decomposition \cite{oseledets2011tensor} can also be utilized to reduce the number of trainable parameters in the tensor $C$:
\begin{equation}\label{eq:TT}
\displaystyle
\begin{aligned}
    c_{\mu, \zeta, z_i, z_j, \beta, \gamma} = \sum_{\alpha_1 = 1}^{r_1} \dots \sum_{\alpha_5 = 1}^{r_5} G^{(1)}_{\zeta, \alpha_1} G^{(2)}_{\alpha_1, z_i, \alpha_2} \\
    \displaystyle
     G^{(3)}_{\alpha_2, \beta, \alpha_3} G^{(4)}_{\alpha_3, z_j, \alpha_4} G^{(5)}_{\alpha_4, \gamma , \alpha_5} G^{(6)}_{\alpha_5, \mu}, 
\end{aligned}
\end{equation}
where five ranks $r_1, \dots, r_5$ should be specified. TT-decomposition cores $G^{(1)}$ and $G^{(6)}$ are matrices, and cores $G^{(2)}, \dots, G^{(5)}$ are third-order tensors. The number of trainable parameters becomes $N_{\text{rad}} r_1 + r_1  N_{\text{spec}} r_2 + r_2 N_{\text{mag}} r_3 + r_3 N_{\text{spec}} r_4 + r_4 N_{\text{mag}} r_5 + r_5 N_{\text{rf}}$.

\subsection{Compressed magnetic Moment Tensor Potential}

\begin{figure*}[!ht]
\begin{center}
\includegraphics[scale=0.2]{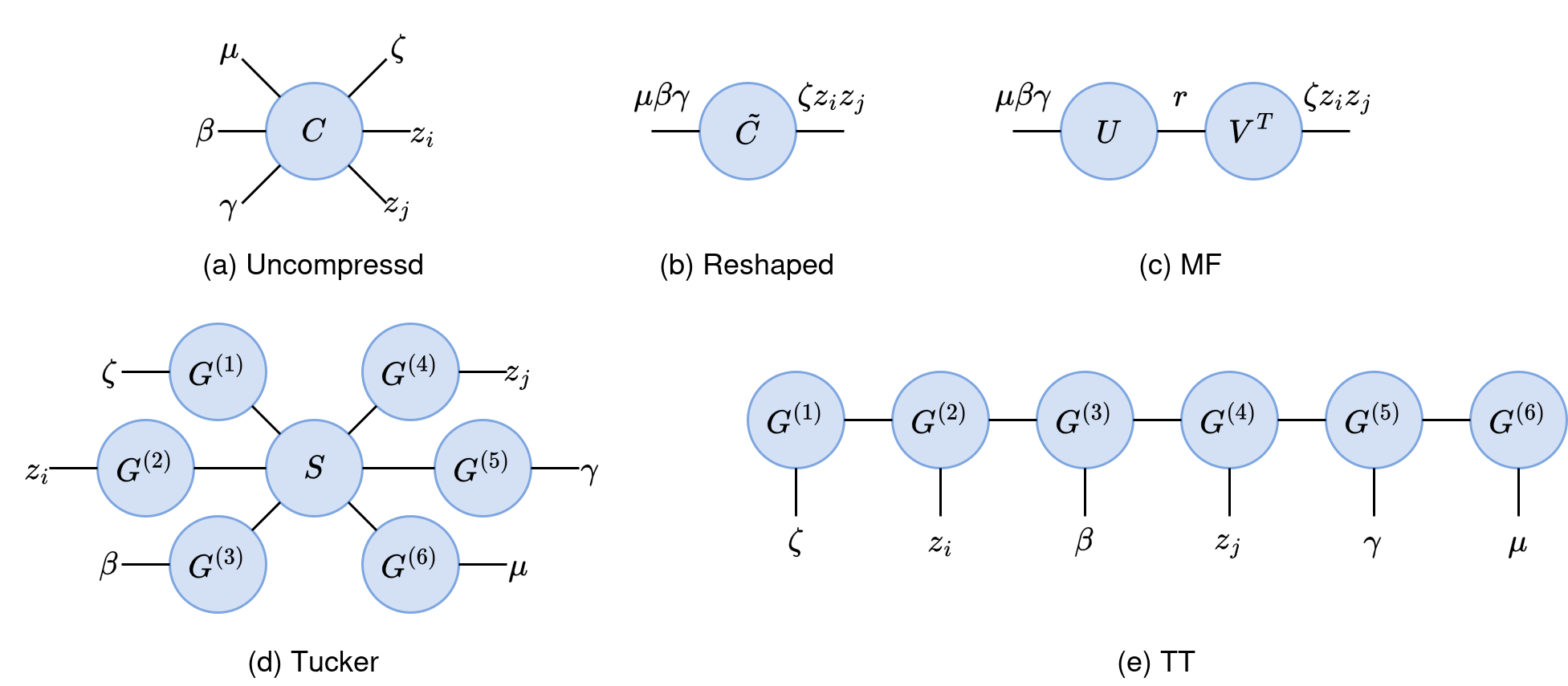}\caption{Representation of used decompositions as tensor diagrams: (a) original uncompressed tensor of radial parameters $C$, (b) $C$ reshaped into matrix $\tilde{C}$, (c) matrix factorization, (d) Tucker decomposition, (e) Tensor Train decomposition.}
\label{fig:TensorDecompositions}
\end{center}
\end{figure*}

The energy calculation in the compressed magnetic Moment Tensor Potential is identical to that in the uncompressed case; the only difference is the decomposition of the tensor $C$ used to construct the radial functions in Eq. \eqref{eq:radial}. In this work, we use three different tensor decompositions: matrix factorization (MF), Tucker decomposition, and Tensor Train decomposition (TT), which were discussed in more detail in the previous section. For MF, we tested three different ways to reshape the original tensor $C$ before factorization:

\begin{equation}\label{eq:MF_variants}
\displaystyle
\begin{array}{c}
    \tilde{C}_1 = \text{reshape}_{(N_{\text{rf}} N_{\text{mag}}^2, N_{\text{rad}} N_{\text{spec}}^2)}(\hat{C}_1), \\
    \displaystyle
    \tilde{C}_2 = \text{reshape}_{(N_{\text{spec}}^2 N_{\text{mag}}^2, N_{\text{rf}} N_{\text{rad}})}(\hat{C}_2),
    \\
    \displaystyle
    \tilde{C}_3 = \text{reshape}_{(N_{\text{rf}} N_{\text{spec}} N_{\text{mag}}, N_{\text{rad}} N_{\text{spec}} N_{\text{mag}})}(\hat{C}_3).
\end{array}
\end{equation}
Here, $\hat{C}_1$, $\hat{C}_2$, and $\hat{C}_3$ correspond to the tensors obtained by permuting the axes of $C$ according to the respective reshaping operations. The resulting matrix factorization variants are referred to as $\text{MF}_1$, $\text{MF}_2$, and $\text{MF}_3$, respectively. For each reshaping, $N_{\text{left}}$ and $N_{\text{right}}$ were chosen such that the resulting matrices $\tilde{C}_{k}$ had dimensions close to square. All decompositions used in this work are presented in the form of tensor diagrams in Figure \ref{fig:TensorDecompositions}.

\subsection{Fitting details}

Training was carried out on two datasets: with Fe--Al configurations from Ref. \cite{kotykhov2024fitting} and with CrN configurations from Ref. \cite{kotykhov2025actively}. For the present work, validation sets were selected from the corresponding configuration pools, comprising 15\% of the total number of configurations. The Fe–Al dataset consisted of 2238 training and 394 validation configurations, while the CrN dataset contained 2060 training and 363 validation configurations. The Fe--Al configurations contain 16 atoms per cell, whereas the CrN configurations have 64 atoms. In all subsequent tests, the radial basis size $N_{\text{rad}}$ was set to 8, and the potential cutoff radius $R_{\text{cut}}$ was set to 4.5 $\text{\AA}$ for the Fe--Al system and 5.0 $\text{\AA}$ for the CrN system.

The optimal mMTP parameters $\Theta$ were determined by minimizing a loss function that includes the squared differences between the predicted and target energies $E$, forces $\textbf{F}$, stresses $\sigma$ (represented as a $3 \times 3$ matrix), and magnetic forces $T$ ($T_i = -\frac{\partial E}{\partial m_i}$):

\begin{equation}\label{eq:loss}
\displaystyle
\begin{array}{c}
    L(\Theta) = \sum_{k=1}^{N_{\text{cfg}}} \Bigr[ w_e (E_k^{\text{MLIP}}(\Theta) - E_k^{\text{DFT}})^2 + 
    \\ 
    \displaystyle
    w_f \sum_{i=1}^{N_{\text{atoms}}(k)} (\textbf{F}_{i, k}^{\text{MLIP}}(\Theta) - \textbf{F}_{i, k}^{\text{DFT}})^2 + \\
    \displaystyle
    w_s \sum_{a,b=1}^{3} (\sigma_{ab, k}^{\text{MLIP}}(\Theta) - \sigma_{ab, k}^{\text{DFT}})^2  +
    \displaystyle
    \\ w_t \sum_{i=1}^{N_{\text{atoms}}(k)} (T_{i, k}^{\text{MLIP}}(\Theta) - T_{i, k}^{\text{DFT}})^2 \Bigl], 
\end{array}
\end{equation}
where $N_{\text{cfg}}$ denotes the number of configurations in the training set, and $N_{\text{atoms}}(k)$ is the number of atoms in configuration $k$. The weights $w_e$, $w_f$ and $w_t$ were chosen to be 1.0, 0.01 $\text{\AA}^2$ and 0.1 $\mu_B^2$ respectively, more details on the choice of these weights can be found in Ref. \cite{kotykhov2024fitting}. The stress weights $w_s$ were set to 0.001 for training on the Fe--Al dataset and to 0.01 for the CrN dataset to improve the accuracy of the calculated elastic constants.

The loss function was minimized using the BFGS algorithm \cite{broyden1970convergence, fletcher1970new, goldfarb1970family, shanno1970conditioning}. The initial values of the trainable parameters are selected randomly from the interval $(-0.3, 0.3)$ for the Fe--Al dataset and from $(-0.1, 0.1)$ for the CrN dataset. To improve the stability of subsequent training, only the parameters $\Xi$ and $X$ were optimized for 25 steps during a pretraining stage before the main training. This pretraining procedure was used for the TT and Tucker models on both the Fe–Al and CrN datasets, whereas it was used for the $\text{MF}_1$ model only on the CrN dataset.

\section{Results}

\subsection{Comparison of different decompositions on example of the Fe--Al system}

\begin{figure*}[!ht]
\begin{center}
\includegraphics[scale=0.39]{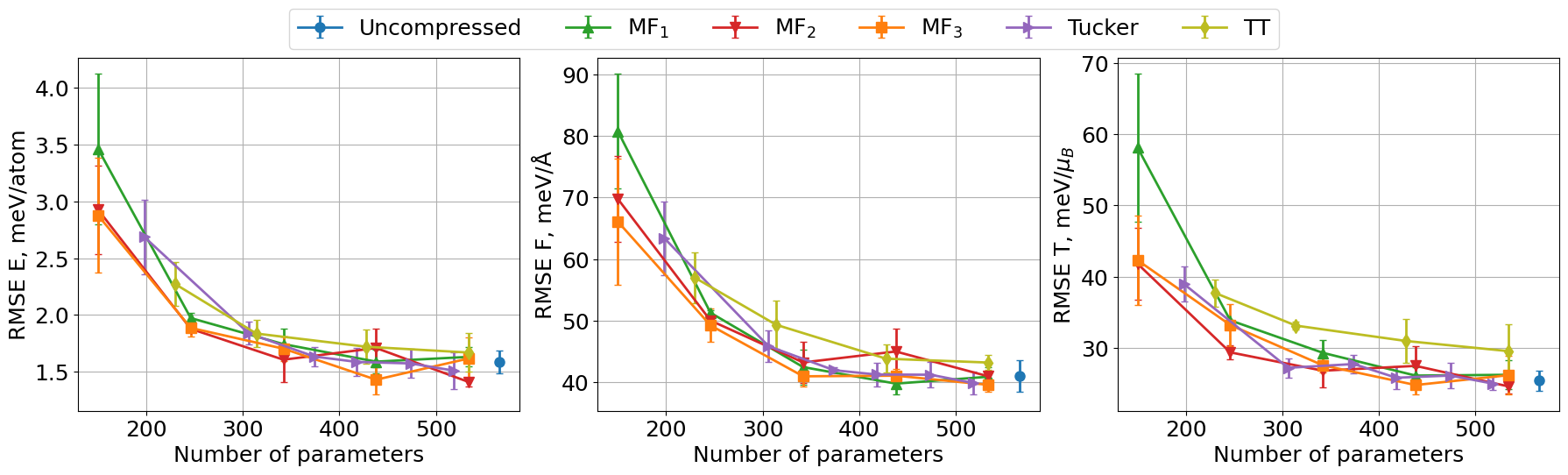}\caption{Average validation root-mean-square errors for energies, forces, and magnetic forces on the Fe--Al dataset for the 14-th level mMTP with magnetic basis $N_{\text{mag}}$ = 2. Results for different variants of matrix factorization (MF), Tucker, and Tensor Train (TT) decompositions are shown. Errors were averaged over an ensemble of potentials. We provide the results with 68\% confidence interval (i.e., 1-$\sigma$ interval). Compressed potentials achieve an approximately $1.5-$fold reduction in the number of parameters while maintaining validation errors comparable to those of uncompressed potentials.}
\label{fig:FeAl_decomp_comparison_14_2}
\end{center}
\end{figure*}

\begin{figure*}[!ht]
\begin{center}
\includegraphics[scale=0.39]{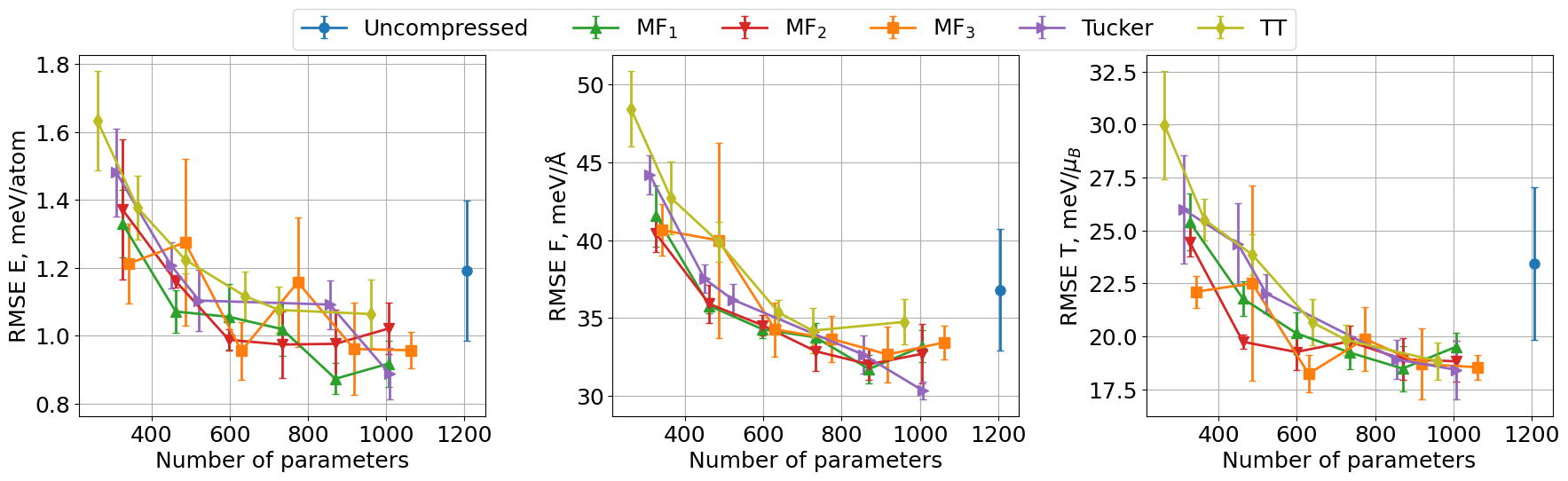}\caption{Average validation root-mean-square errors for energies, forces, and magnetic forces on the Fe--Al dataset for the 14-th level mMTP with magnetic basis $N_{\text{mag}}$ = 3. Results for different variants of matrix factorization (MF), Tucker, and Tensor Train (TT) decompositions are shown. Errors were averaged over an ensemble of potentials. We provide the results with 68\% confidence interval (i.e., 1-$\sigma$ interval). Compressed potentials achieve an approximately 3-fold reduction in the number of parameters while maintaining validation errors comparable to those of uncompressed potentials.}
\label{fig:FeAl_decomp_comparison_14_3}
\end{center}
\end{figure*}

We first evaluated various tensor decompositions in terms of the trade-off between compression and predictive performance on a validation set for the Fe--Al system. For each set of hyperparameters, an ensemble of five potentials was trained. A potential was excluded from the ensemble and from subsequent property calculations if its training loss exceeded 2.5 times the minimum training loss observed among all potentials in the ensemble. The average root-mean-square validation errors for energies, forces, and magnetic forces for the 14-th level potential with magnetic basis size $N_{\text{mag}}$ = 2 and $N_{\text{mag}}$ = 3 are shown in Figures \ref{fig:FeAl_decomp_comparison_14_2} and \ref{fig:FeAl_decomp_comparison_14_3}, respectively. For all compression methods, a characteristic trend is observed: at moderate compression ratios, the model’s predictive performance sometimes improves relative to the uncompressed model, accompanied by greater training stability, as evidenced by a smaller spread of errors. At high compression ratios, however, validation errors increase as the number of model parameters decreases. Here, the compression ratio is defined as the ratio between the numbers of parameters in the uncompressed and compressed models. For two basis magnetic functions, all decomposition schemes, except for TT, which exhibits higher errors, provide roughly comparable compression while maintaining errors at the level of the uncompressed model. However, for three magnetic basis functions, $\text{MF}_1$, $\text{MF}_2$, and Tucker achieve the most favorable compression-to-error ratios, whereas $\text{MF}_3$ exhibits substantially larger variability, as reflected by its wider error bars.

For further testing, the $\text{MF}_1$ scheme was identified as the most efficient scheme for compression, because it provides one of the best reductions in the number of parameters while maintaining relatively stable training and requires only one additional hyperparameter $\{r\}$, compared to six hyperparameters $\{r_1, r_2, r_3, r_4, r_5, r_6\}$ required by the Tucker decomposition.

\subsection{Compression efficiency for different potential hyperparameters}\label{sec:FeAl_hypers_test}

\begin{figure}[!ht]
\begin{center}
\includegraphics[scale=0.35]{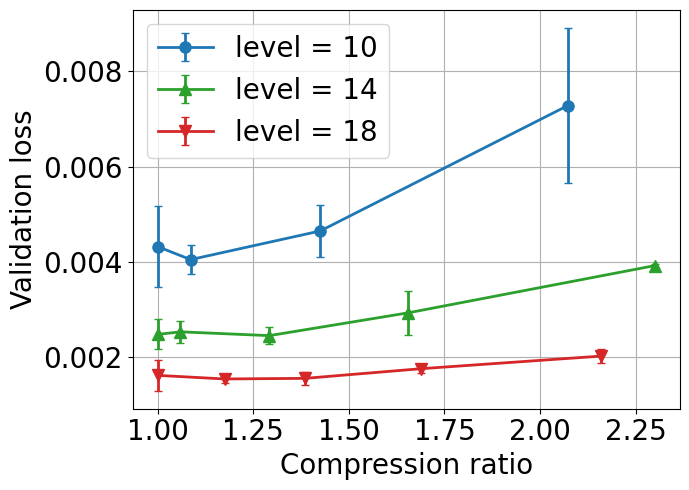}\caption{Dependence of the validation loss on the mMTP compression ratio for the Fe--Al dataset at different potential levels and a fixed magnetic basis size of $N_{\text{mag}}$ = 2. The $\text{MF}_1$ scheme was used for potential compression. A compression ratio of 1.0 corresponds to an uncompressed model. A noticeable increase in the maximum achievable compression ratio, at which the loss of the compressed model does not exceed that of the uncompressed model, is observed as the potential level increases. Averaging was performed over an ensemble of potentials. We provide the results with 68\% confidence interval (i.e., 1-$\sigma$ interval).}
\label{fig:FeAl_valid_loss_from_level}
\end{center}
\end{figure}

\begin{figure}[!ht]
\begin{center}
\includegraphics[scale=0.35]{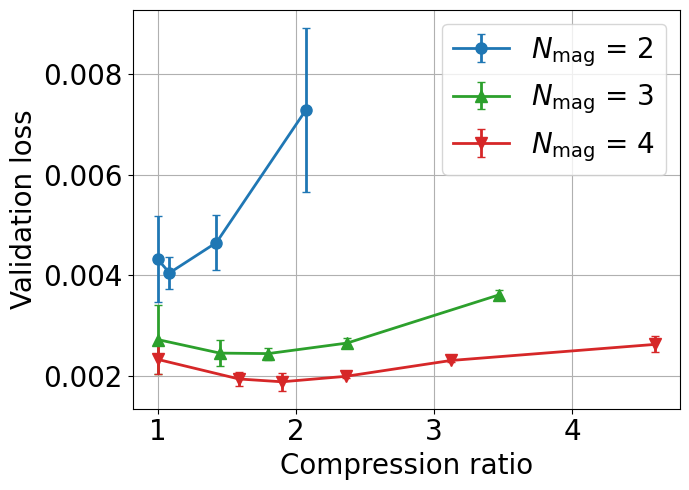}\caption{Dependence of the validation loss on the mMTP compression ratio  for the Fe--Al dataset at different magnetic basis sizes and a fixed potential level = 10. The $\text{MF}_1$ scheme was used for potential compression. A compression ratio of 1.0 corresponds to an uncompressed model. A significant increase in the maximum achievable compression ratio, at which the loss of the compressed model does not exceed that of the uncompressed model, is observed as the magnetic basis size increases. Averaging was performed over an ensemble of potentials. We provide the results with 68\% confidence interval (i.e., 1-$\sigma$ interval)..}
\label{fig:FeAl_valid_loss_from_N_mag}
\end{center}
\end{figure}

This section examines the compression efficiency, defined as the maximum achievable compression ratio for which the validation loss does not exceed that of the uncompressed model, for two key mMTP hyperparameters: the potential level and the magnetic basis size $N_{\text{mag}}$. Training was performed on the Fe--Al system using the $\text{MF}_1$ compression scheme with varying rank ${r}$. The dependence of the validation loss on the compression ratio for a fixed $N_{\text{mag}} = 2$ and different potential levels is shown in Figure \ref{fig:FeAl_valid_loss_from_level}, while the corresponding dependence for different magnetic basis sizes at a fixed potential level of 10 is shown in Figure \ref{fig:FeAl_valid_loss_from_N_mag}.

The compression efficiency of mMTP increases with both the potential level and the magnetic basis size. Moreover, for large magnetic basis sets, mMTP compression is more efficient than for larger potential levels. For example,  compression efficiency rises from approximately 1.25 to 3 when the magnetic basis size is increased from 2 to 4. Furthermore, at moderate compression ratios, the average validation loss is slightly lower than that of the uncompressed model, while the spread of the results is also reduced. This may indicate that the compressed models are trained more stably or efficiently.

\begin{figure}[!ht]
\begin{center}
\includegraphics[scale=0.35]{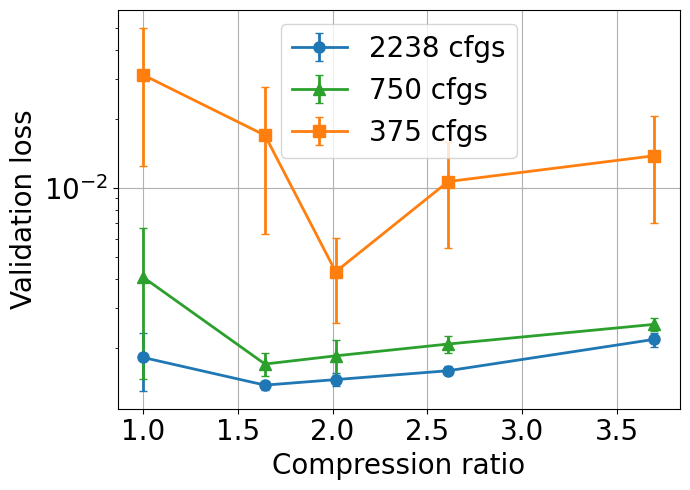}\caption{Dependence of the validation loss on the mMTP compression ratio for different training set sizes of the Fe--Al dataset. Potentials of the 14-th level with $N_{\text{mag}} = 3$ were used. The $\text{MF}_1$ scheme was used for potential compression. A compression ratio of 1.0 corresponds to an uncompressed model. It is noticeable that even with fewer configurations in the training sample, the compression efficiency of mMTP does not deteriorate. The validation loss is shown on a logarithmic scale. Averaging was performed over an ensemble of potentials. We provide the results with 68\% confidence interval (i.e., 1-$\sigma$ interval).}
\label{fig:FeAl_valid_loss_from_train_size}
\end{center}
\end{figure}

The effectiveness of potential compression was also evaluated for different training set sizes. To that end, the original training set of 2238 configurations was randomly reduced to 750 and 375 configurations, while the validation set was kept unchanged. The dependence of the validation loss on the mMTP compression ratio for different training set sizes is presented in Figure \ref{fig:FeAl_valid_loss_from_train_size}. A 14-th level potential with $N_{\text{mag}}=3$ was used for this test. Here, potentials whose validation loss exceeded the minimum ensemble loss by a factor of 2.5 were considered unsuccessfully converged and were therefore excluded from the ensemble and subsequent property calculations. Even with a reduced training set, the mMTP potentials can still be effectively compressed. For example, for a training set of 750 configurations, compression can even improve the stability of the potential, as indicated by the smaller spread of validation losses for the compressed models.

\subsection{Properties of the Fe--Al system predicted with compressed mMTPs}

To assess the performance of compressed mMTPs, the formation energies of various configurations, as well as the dependence of the lattice parameters and cell magnetic moment on the Al concentration, were calculated for the Fe–Al system. To that end, geometric optimization was performed for 21 configurations containing 16 atoms per cell, with varying concentrations and arrangements of Al atoms. More details on the configurations used can be found in \cite{friak2010ab, kotykhov2023constrained}. It should be noted that after each geometric optimization step, the equilibrium magnetic moments were determined for the updated atomic positions and lattice parameters. The ASE package \cite{hjorth2017atomic} was used for geometry optimization. 

The root-mean-square errors of the predictions of the formation energies of configurations, equilibrium magnetic moments, and averaged lattice parameters for the different compressed mMTPs of the 14-th level with $N_{\text{mag}} = 3$ are presented in Table \ref{tab:rmses}. Moderate compression ratios between 1.5 and 2 do not lead to a loss of predictive accuracy. At these compression ratios, the modeling errors remain comparable to, and in some cases are marginally lower than those of the uncompressed model. With few exceptions, the modeling errors increase monotonically with the compression ratio. Among the compression methods considered, matrix factorization (MF) and Tucker decomposition yield the lowest root-mean-square errors at comparable compression ratios.

\begin{table*}
\caption{Root-mean-square errors of prediction of the formation energy ($E_{\text{form}}$), equilibrium magnetic moments ($M_{\text{equil}}$), and averaged lattice parameters ($a$) for 21 considered configurations of the Fe–Al system using various compressed mMTPs. Potentials of the 14-th level with $N_{\text{mag}} = 3$ were used. Averaging was performed over an ensemble of potentials. We provide the results with 68\% confidence interval (i.e., 1-$\sigma$ interval).}
\label{tab:rmses}
\begin{center}
\begin{tabular}{cccccc}
\hline
Model & Ranks      & Compression & $E_{\text{form}}$, meV/atom & $M_{\text{equil}}$, $\mu_b$ & $a$, Å \\ \hline
Uncompressed           & -                    & 1              & 0.46±0.03                & 0.039±0.006            & 0.0016±0.0005 \\ \hline
$\text{MF}_1$            & \{10\}               & 1.64           & 0.46±0.10                & 0.033±0.003            & 0.0013±0.0001 \\
$\text{MF}_1$            & \{8\}                & 2.02           & 0.48±0.10                & 0.030±0.003            & 0.0013±0.0001 \\
$\text{MF}_1$            & \{6\}                & 2.61           & 0.51±0.06                & 0.037±0.007            & 0.0016±0.0002 \\
$\text{MF}_1$            & \{4\}                & 3.7            & 0.46±0.07                & 0.035±0.008            & 0.0016±0.0002 \\ \hline
Tucker         & \{5, 2, 3, 2, 3, 4\} & 1.41           & 0.41±0.05                & 0.032±0.002            & 0.0013±0.0001 \\
Tucker         & \{6, 2, 2, 2, 2, 4\} & 2.31           & 0.52±0.08                & 0.038±0.006            & 0.0016±0.0003 \\
Tucker         & \{5, 2, 2, 2, 2, 4\} & 2.68           & 0.57±0.13                & 0.044±0.008            & 0.0017±0.0003 \\
Tucker         & \{4, 2, 2, 2, 2, 3\} & 3.89           & 0.57±0.02                & 0.043±0.004            & 0.0020±0.0002 \\ \hline
TT             & \{5, 7, 10, 7, 5\}   & 1.89           & 0.49±0.03                & 0.034±0.003            & 0.0014±0.0001 \\
TT             & \{6, 6, 6, 6, 6\}    & 2.48           & 0.72±0.18                & 0.040±0.003            & 0.0017±0.0001 \\
TT             & \{5, 5, 5, 5, 5\}    & 3.31           & 0.59±0.07                & 0.041±0.009            & 0.0019±0.0002 \\ \hline
\end{tabular}
\end{center}
\end{table*}

\begin{figure}[!ht]
\begin{center}
\includegraphics[scale=0.40]{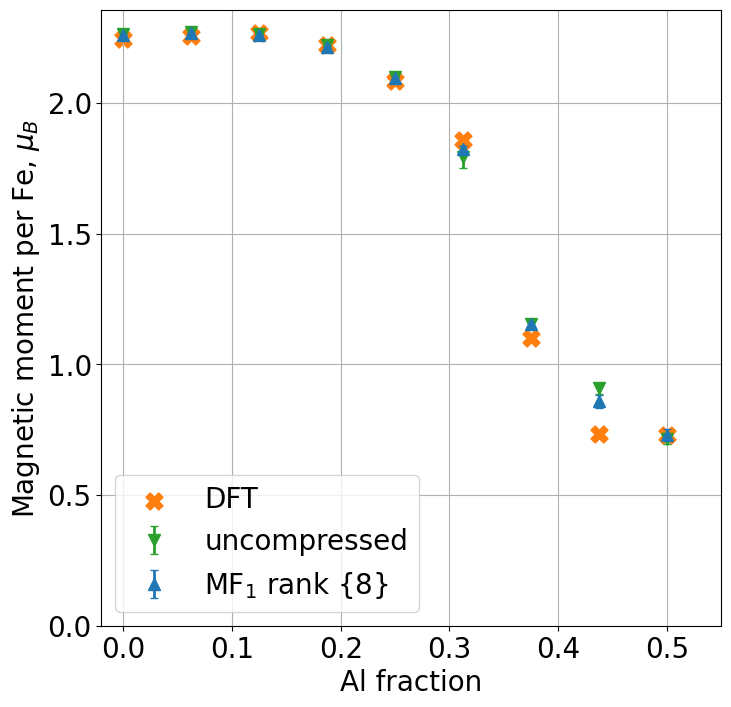}\caption{Total magnetic moment per Fe atom as a function of Al concentration for the Fe–Al system, obtained using uncompressed and compressed ($\text{MF}_1$ with $r=8$, compression ratio $\approx$ 2.0) mMTP potentials. Potentials of the 14-th level and with $N_{\text{mag}} = 3$ were used. Averaging was performed over an ensemble of potentials. We provide the results with 68\% confidence interval (i.e., 1-$\sigma$ interval).}
\label{fig:FeAl_magnetization}
\end{center}
\end{figure}

Figure \ref{fig:FeAl_magnetization} shows the total cell magnetic moment per Fe atom as a function of Al concentration for the uncompressed and compressed ($\text{MF}_1$ with $r=8$, compression ratio $\approx$ 2.0) mMTP, considering only configurations on the convex hull. The compressed model closely reproduces both the uncompressed model and the DFT reference.

\begin{figure}[!ht]
\begin{center}
\includegraphics[scale=0.40]{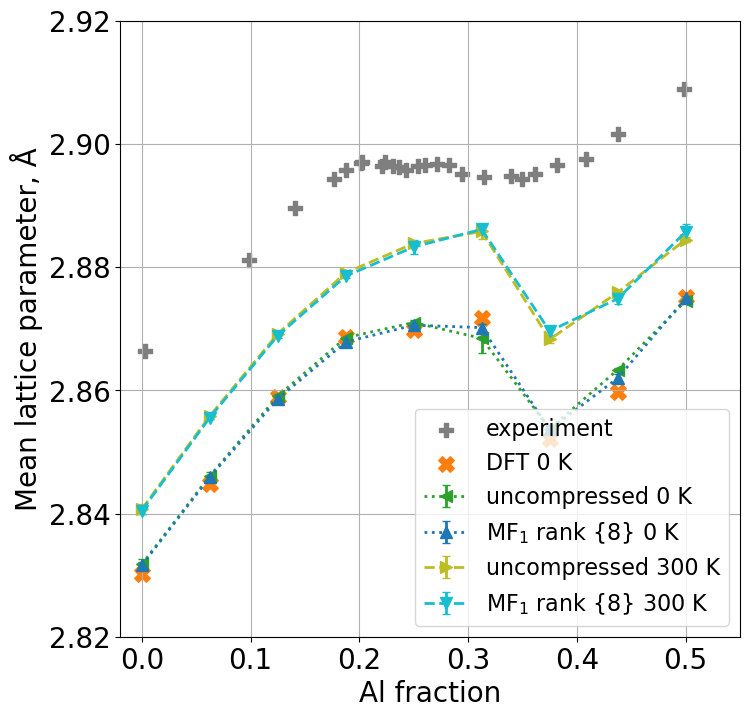}\caption{Average lattice parameter as a function of Al concentration for the Fe–Al system, obtained using uncompressed and compressed ($\text{MF}_1$ with $r=8$, compression ratio $\approx$ 2.0) mMTP potentials. Potentials of the 14-th level and with $N_{\text{mag}}$ = 3 were used. Averaging was performed over an ensemble of potentials. We provide the results with 68\% confidence interval (i.e., 1-$\sigma$ interval). Experimental data are taken from \cite{taylor1958constitution}.}
\label{fig:FeAl_lattice}
\end{center}
\end{figure}

The dependence of the lattice parameter on Al concentration is shown in Figure \ref{fig:FeAl_lattice} for ensembles of uncompressed and compressed mMTPs, using the same compression settings as described above. The data at 300 K were obtained from NPT molecular dynamics for 50000 steps using the LAMMPS package \cite{LAMMPS}. Trajectories that resulted in a simulation error, such as the loss of atoms or non-numeric pressure, were excluded from the averaging. The results obtained with the compressed model are numerically close to those of both the DFT reference and the uncompressed model. Moreover, the compressed model exhibits a higher fraction of converged trajectories (89\% versus 80\%). Both the compressed and uncompressed mMTPs qualitatively reproduce the experimental dependence of the lattice parameter on Al concentration \cite{taylor1958constitution}. However, a noticeable quantitative underestimation is observed, which is likely attributable to the underestimation of the lattice parameters of bcc Fe and bcc Fe–Al at the DFT level \cite{dal2000ab, friak2010ab}.

\subsection{Modeling with compressed mMTP on example of the CrN system}

\begin{figure}[!ht]
\begin{center}
\includegraphics[scale=0.40]{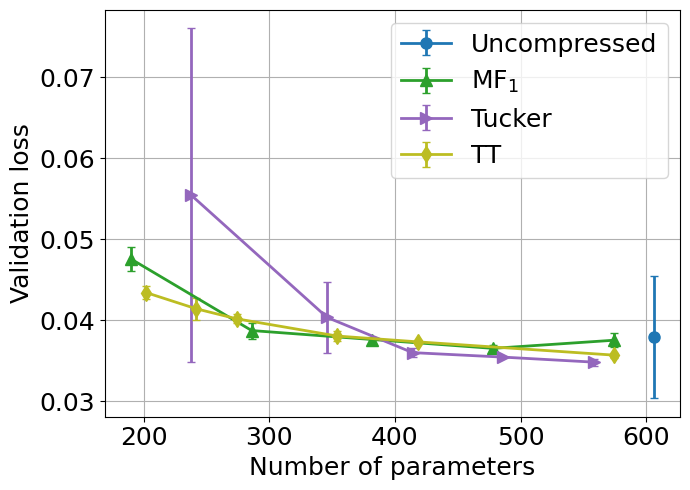}\caption{Validation loss as a function of the number of trainable parameters for the CrN dataset, comparing the uncompressed mMTP with its compressed variants ($\text{MF}_1$, Tucker, TT). Potentials of the 16-th level and with $N_{\text{mag}} = 2$ were used. Averaging was performed over an ensemble of potentials. We provide the results with 68\% confidence interval (i.e., 1-$\sigma$ interval). The large spread of values for Tucker decomposition at 238 parameters times is caused by a single underfitted potential with a validation loss of 0.08.}
\label{fig:CrN_valid_loss}
\end{center}
\end{figure}

In this section, we consider the results of modeling the B1 phase of CrN with compressed mMTPs. All calculations were performed using potentials of level 16 with $N_{\text{mag}} = 2$.
 Figure \ref{fig:CrN_valid_loss} shows the dependence of the validation loss on the number of trainable parameters for different ranks and compression schemes. Similar to the Fe--Al system, the mMTP for the CrN system can be compressed by a factor of 1.5–2 without loss of predictive performance on the validation set. The $\text{MF}_1$, Tucker, and TT compression schemes provide comparable results for parameter compressions of up to approximately $1.5-$fold. However, at higher compression ratios, the Tucker decomposition exhibits a larger spread of values. We also emphasize that the convergence behavior varied significantly across decompositions. The Tucker decomposition converged for only approximately half of the ensemble, whereas TT achieved a convergence rate of 0.86. In contrast, $\text{MF}_1$ converged for all five potentials. Due to a higher fraction of converged models and a smaller number of ranks required, a matrix factorization approach was selected for further modeling. 

\begin{figure}[!ht]
\begin{center}
\includegraphics[scale=0.40]{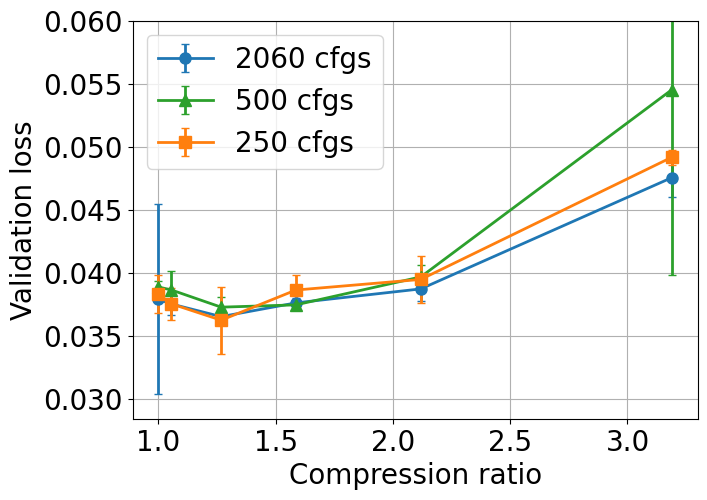}\caption{Dependence of the validation loss on the mMTP compression ratio for different training size of the CrN dataset. Potentials of the 16-th level and with $N_{\text{mag}} = 2$ were used. The $\text{MF}_1$ scheme was used for potential compression. A compression ratio of 1.0 corresponds to an uncompressed model. Averaging was performed over an ensemble of potentials. We provide the results with 68\% confidence interval (i.e., 1-$\sigma$ interval). The large spread of values for 500 configurations and a compression ratio of 3.2 times is caused by a single underfitted potential with a validation loss of 0.08.}
\label{fig:CrN_valid_loss_from_train_size}
\end{center}
\end{figure}

Following the same procedure as in the section \ref{sec:FeAl_hypers_test}, we tested compressed mMTPs by reducing the training set size and comparing the validation performance of the $\text{MF}_1$‑compressed potential with that of the uncompressed potential. The results are presented in Figure \ref{fig:CrN_valid_loss_from_train_size}. For both the full training set (2060 configurations) and the reduced sets (500 and 250 configurations), a similar dependence of the validation loss on the compression ratio is observed. Thus, as in the Fe--Al system, compressed potentials remain effective for different training set sizes at moderate compression ratios.

The paramagnetic state was modeled by averaging forces and stresses over 50 random magnetic configurations, following the approach used in the previous study with the uncompressed potential \cite{kotykhov2025actively}. Specifically, for a given set of atomic coordinates and atom types, the magnetic moment directions of Cr atoms were assigned randomly, with half of the moments oriented upward and half downward. The moment amplitudes were then equilibrated, after which the forces and stresses were averaged. The elastic constants of paramagnetic CrN obtained using compressed and uncompressed mMTPs are presented in Table \ref{tab:CrN_elasctic}, together with a comparison to first-principles calculations from Ref. \cite{zhou2014structural}. The elastic constants predicted by the compressed potentials are numerically close to those obtained with the uncompressed potential and to the reference DFT values.

\begin{table}
\caption{Elastic constants ($\text{C}_{ij}$) of paramagnetic CrN obtained using different compression ratios (Comp.). Potentials of the 16-th level with $N_{\text{mag}} = 2$ were used. Averaging was performed over an ensemble of
potentials. We provide the results with 95\% confidence interval (i.e., 2-$\sigma$ interval).}
\label{tab:CrN_elasctic}
\begin{tabular}{ccccc}
\hline
Model        & Comp. & $\text{C}_{11}$         & $\text{C}_{12}$         & $\text{C}_{44}$          \\ \hline
Uncompressed & 1.0            & 657 ± 6 & 93 ± 13 & 150 ± 4  \\ \hline
$\text{MF}_1$ \{6\}    & 1.6          & 663 ± 8 & 92 ± 5  & 147 ± 11 \\
$\text{MF}_1$ \{4\}    & 2.1          & 659 ± 2 & 98 ± 3  & 146 ± 10 \\ \hline
DFT from \cite{zhou2014structural}          & -            & 649         & 99          & 145          \\ \hline
\end{tabular}
\end{table}

\begin{figure}[!ht]
\begin{center}
\includegraphics[scale=0.30]{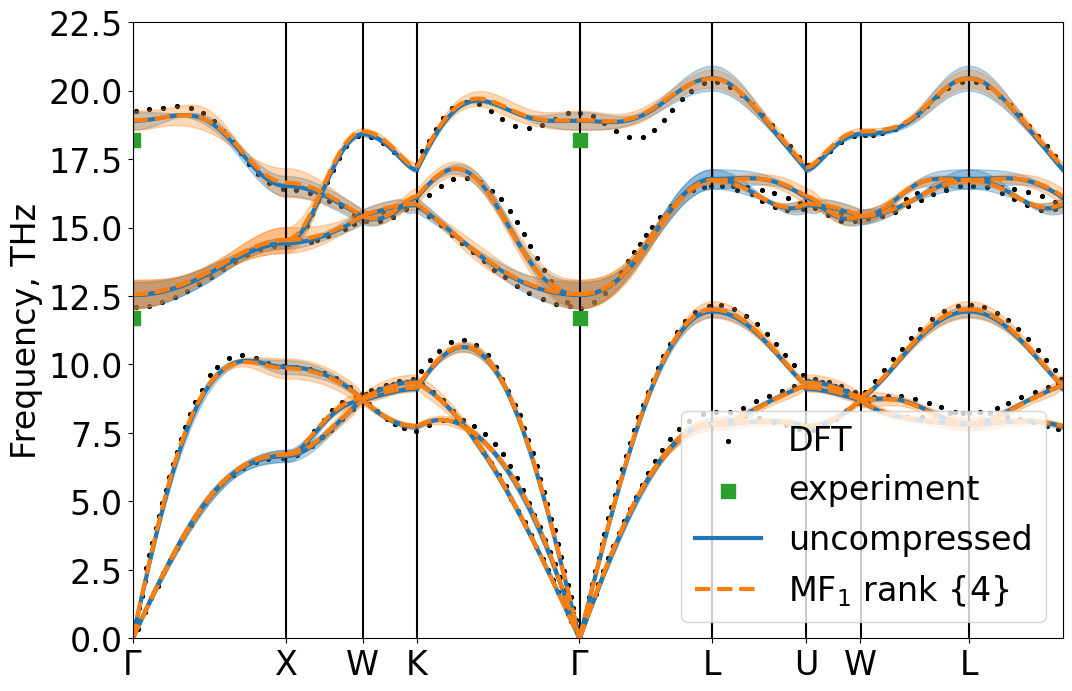}\caption{Phonon spectrum of paramagnetic CrN obtained using uncompressed and compressed ($\text{MF}_1$ with $r=4$, compression ratio $\approx$ 2.1) mMTPs. Potentials of the 16-th level with $N_{\text{mag}} = 2$ were used. Averaging was performed over an ensemble of potentials. We provide the results with 99.7\% confidence interval (i.e., 3-$\sigma$ interval). The reference DFT data are taken from Ref. \cite{zhou2014structural}, and experimental data from Ref. \cite{zhang2010crn}.}
\label{fig:CrN_phonons}
\end{center}
\end{figure}

The phonon dispersion curves of paramagnetic CrN for the compressed ($\text{MF}_1$ with $r=4$, compression ratio $\approx$ 2.1) and uncompressed models, together with experimental data from Ref. \cite{zhang2010crn} and first-principles data from Ref. \cite{zhou2014structural} are shown in Figure \ref{fig:CrN_phonons}. The phonon spectra were calculated using the Phonopy package \cite{togo2023implementation, togo2023first} with non-analytic term corrections to the dynamic matrix \cite{wang2010mixed}. The experimental values from \cite{zhang2010crn} were used for the dielectric tensor and the Born effective charge matrix. The predicted dispersion curves for the uncompressed and approximately two-fold compressed models are in good agreement.

\section{Conclusion}

In this study, we proposed and evaluated several tensor decomposition approaches based on matrix factorization, Tucker decomposition, and Tensor Train decomposition for the radial parameters of magnetic Moment Tensor Potential (mMTP). The developed compressed models enable reducing the number of trainable parameters by a factor of 1.5–3 with respect to the uncompressed mMTP, depending on the selected potential hyperparameters, while maintaining a comparable validation error. A comparison of the different compression schemes showed that matrix factorization is the most optimal in terms of validation errors and the number of ranks required for the compressed model. The proposed compressed mMTP models were also tested in simulations for the Fe--Al and CrN systems. For the Fe--Al system, the dependencies of the cell magnetic moment and lattice parameter on the Al concentration were calculated, and molecular dynamics simulations were performed at a temperature of 300 K. The results obtained with the compressed and uncompressed models were in close numerical agreement and were also consistent with reference first-principles calculations. For the CrN system, the elastic constants and phonon spectrum for the paramagnetic state were calculated. In this case, good agreement was likewise observed between the compressed and uncompressed models, as well as between the calculated results, available density functional theory (DFT) data, and experimental measurements. The compression efficiency was evaluated for a range of mMTP hyperparameters. The maximum compression ratio that can be achieved without loss of validation quality increases with the magnetic basis size and potential level.

In the future, compressed magnetic models can be used for active learning of potentials, and the reduction in the number of trainable parameters is expected to reduce the amount of selected configurations during active learning. Further improvements may also be achieved by employing more advanced decomposition techniques, such as adaptive tensor networks \cite{nie2021adaptive}, which can simplify the selection of decomposition ranks and axis orderings by avoiding manual hyperparameter tuning.

\section{CRediT authorship contribution statement}

{\bf Alexey S. Kotykhov}: Software, Validation, Investigation, Data curation, Writing - Original Draft, Visualization.
{\bf Alexander V. Shapeev}: Conceptualization, Writing - Review \& Editing.
{\bf Maxim Rakhuba}: Conceptualization, Methodology, Formal analysis, Writing - Review \& Editing
{\bf Ivan S. Novikov}: Conceptualization, Methodology, Formal analysis, Resources, Writing - Review \& Editing, Supervision, Funding acquisition.

\section{Declaration of competing interest}

The authors declare that they have no known competing financial interests or personal relationships that could have appeared to influence the work reported in this paper.

\section{Acknowledgments}

This work was supported by Russian Science Foundation (grant number 26-73-10173, https://rscf.ru/project/26-73-10173/). The supercomputer of HSE University was used to conduct the computational research \cite{kostenetskiy2021hpc}.

\section{Data availability}

Data will be made available on request.

\bibliographystyle{elsarticle-num}

\end{document}